\documentclass[aps,prx,twocolumn,nofootinbib,groupedaddress,superscriptaddress,longbibliography]{revtex4-2}
\usepackage{amssymb}
\usepackage{multirow}
\usepackage{graphicx}
\usepackage{amsmath}
\usepackage{color}
\usepackage{mathrsfs}
\usepackage{float}
\usepackage{indentfirst}
\usepackage{mathrsfs}
\usepackage{float}
\usepackage{indentfirst}
\usepackage{textcomp}
\usepackage{comment}
\usepackage{mathtools}
\usepackage{natbib,hyperref}
\usepackage{soul}

\begin{document}
\title{On the correlation between entanglement and the negative sign problem}
%%%%%%%%%%%%

%%%%%%%%%%%%
\author{Ping Xu}
\affiliation{Key Laboratory of Artificial Structures and Quantum Control (Ministry of Education),  School of Physics and Astronomy, Shanghai Jiao Tong University, Shanghai 200240, China}

\author{Yang Shen}
\affiliation{Key Laboratory of Artificial Structures and Quantum Control (Ministry of Education),  School of Physics and Astronomy, Shanghai Jiao Tong University, Shanghai 200240, China}

\author{Yuan-Yao He}
\affiliation{Institute of Modern Physics, Northwest University, Xi'an 710127, China}
\affiliation{Shaanxi Key Laboratory for Theoretical Physics Frontiers, Xi'an 710127, China}
\affiliation{Hefei National Laboratory, Hefei 230088, China}

\author{Mingpu Qin} \thanks{qinmingpu@sjtu.edu.cn}
\affiliation{Key Laboratory of Artificial Structures and Quantum Control (Ministry of Education),  School of Physics and Astronomy, Shanghai Jiao Tong University, Shanghai 200240, China}

\affiliation{Hefei National Laboratory, Hefei 230088, China}

\date{\today}

%%%%%%%%%%%%

%%%%%%%%%%%%
\begin{abstract}
In this work, we study the correlation between entanglement and the negative sign problem in quantum Monte Carlo for the simulation of low-dimensional strongly correlated quantum many body systems. Entanglement entropy characterizes the difficulty of many-body simulation with tensor network state related methods, while the average sign measures the difficulty in many-body simulation for a variety of quantum Monte Carlo methods. Although there exist cases where one type of method works better than the other, it is desirable to find the possible correlation between entanglement and average sign for general hard strongly correlated systems regarding computational complexity. We take the doped two-dimensional Hubbard model as an example and numerically calculate the doping evolution of both the entanglement in the ground state with Density Matrix Renormalization Group and the average sign in the Auxiliary Field Quantum Monte Carlo simulation at low temperature. The results show that they are indeed correlated. The entanglement entropy (average sign) shows a peak (dip) around $20\%$ doping, indicating that it is the difficult region for both methods. The vicinity of $20\%$ doping is also the most intriguing region in both the Hubbard model and cuprate high-Tc superconductors where competing states with close energy intertwine with each other. Recognizing the correlation between entanglement and average sign provides new insight into our understanding of the difficulty in the simulation of strongly correlated quantum many-body systems. 
\end{abstract}
%%%%%%%%%%%%

%%%%%%%%%%%%
\maketitle

\section{Introduction}

The understanding of exotic phases in strongly correlated quantum many-body systems is one of the most important topics in condensed matter physics \cite{dagottoCorrelatedElectronsHightemperature1994}. The study of these systems now relies mainly on numerical many-body methods because it is difficult to handle the strong correlation effect analytically \cite{leblancSolutionsTwoDimensionalHubbard2015,PhysRevX.11.011058}. The Hubbard model \cite{Hubbardmodel1963} is arguably the simplest strongly correlated electronic model and in two dimension it is believed to be related to cuprate high-Tc superconductors \cite{qinHubbardModelComputational2022,arovasHubbardModel2022}. Over decades of studies, it seems that the two-dimensional (2D) Hubbard model with intermediate to large $U$ and close to half-filling is the most challenging region for existing numerical methods \cite{leblancSolutionsTwoDimensionalHubbard2015}. Experimentally, this is also the most intriguing region where different states close in energy intertwine with each other \cite{RevModPhys.87.457}.
 
These facts make us to ask whether there is a correlation among the quantities which measure the difficulty in the many-body simulation with different many-body methods regarding computational complexity. In this work, we focus on two categories of methods, i.e., tensor network state related methods and quantum Monte Carlo (QMC). For tensor network state related methods \cite{whiteDensityMatrixFormulation1992,schollwockDensitymatrixRenormalizationGroup2011,orusTensorNetworksComplex2019,RevModPhys.93.045003,annurev-conmatphys-040721-022705,xiang2023density}, the entanglement in the ground state of the studied system determines the accuracy one can reach with given resources, while for a variety of QMC methods, the average sign is the measure of the hardness in simulating a quantum many-body system \cite{lohSignProblemNumerical1990a}. In this work, we try to study the correlation between entanglement in the ground state and average sign in the QMC simulation at low temperature quantitatively, by taking the doped 2D Hubbard model as an example. 

At the first glance, it seems that entanglement has nothing to do with the average sign from their definitions. However, previous studies suggest that there indeed exists correlation between them. It is known that the anti-ferromagnetic Heisenberg model on bipartite lattices (e.g., one-dimensional, square, honeycomb lattices) is sign problem free. At the same time, these systems are easier to study with tensor network state related methods than the Heisenberg model on frustrated lattices such as triangular \cite{whiteNeelOrderSquare2007} and kagome \cite{evenblyFrustratedAntiferromagnetsEntanglement2010,yanSpinLiquidGroundState2011,PhysRevLett.118.137202} lattices, which encounter negative sign problem in QMC. In \cite{groverEntanglementSignStructure2015}, it is found that a typical random positive wave-function (which is in principle sign problem free) can't support volume law entanglement. The diffusion Monte Carlo (DMC) method can accurately simulate one-dimensional (1D) fermion system because the node structure of 1D fermion ground state is solely determined by the minus sign from the anti-commutation relation of fermion \cite{ceperleyFermionNodes1991}. Interestingly, it is well known that Density Matrix Renormalization Group (DMRG) \cite{whiteDensityMatrixFormulation1992} can easily handle 1D quantum system because the entanglement entropy in the ground state of 1D system scales at most logarithmically with the size of the system. However, both DMC and DMRG encounter difficulty in the study of 2D systems. On the one hand, the sign structure of 2D quantum system is far more complicated than the 1D case, making it almost impossible to determine the exact sign structure of general system \cite{ceperleyFermionNodes1991,PhysRevB.78.035104,PhysRevB.95.155102}. This can be viewed as a manifestation of the negative sign problem. On the other hand, area law \cite{eisertColloquiumAreaLaws2010} tells that the entanglement entropy of the ground state of 2D systems scales at least as the linear dimension of the systems, making the study of 2D quantum system with DMRG difficult. We notice that there also exist 2D tensor network state methods \cite{orusTensorNetworksComplex2019,RevModPhys.93.045003,annurev-conmatphys-040721-022705,xiang2023density}, but the heavy cost makes the entanglement they can handle limited. In \cite{grossmanRobustFermiLiquidInstabilities2023}, it is found that sign problem-free model can't have a stable Fermi surface, while it is known that the Fermi system with non-trivial Fermi surface usually causes a logarithm correlation to the area law of entanglement \cite{swingleEntanglementEntropyFermi2010a,PhysRevLett.96.010404}. This means that system with negative sign problem in QMC usually has large entanglement in the ground state. In \cite{mondainiQuantumCriticalPoints2022,PhysRevB.107.245144}, it is also shown that sign problem is linked to quantum critical behavior, and it is known that the entanglement usually increases when approaching quantum critical point \cite{vidalEntanglementQuantumCritical2003,zhangEntanglementEntropyCritical2011}. Recent study also shows that the positive definiteness of the elements of tensor network makes the contraction of it easier \cite{PRXQuantum.6.010312}. 
We also notice that the average sign is a contribution term in the second order Renyi entropy defined with the replica trick \cite{PhysRevLett.104.157201,broeckerRenyiEntropiesInteracting2014,broeckerEntanglementFermionSign2016}.

In this work, we take the Hubbard model with size $4 \times  16$ and $U/t =8$ as an example to study the correlation between entanglement and average sign. Our focus is the ground state or low temperature simulations. We carefully study the doping evolution of the entanglement in the ground state with DMRG and the average sign in QMC at low temperature. We find that they are indeed correlated. The entanglement (average sign) shows a peak (dip) around $20 \%$ doping, which means that it is the most difficult region for both QMC and tensor network state related methods. We also notice that the vicinity of $20\%$ doping is the most intriguing region in both Hubbard model and cuprates where competing states with close energies intertwine with each other \cite{RevModPhys.87.457,PhysRevX.10.031016,doi:10.1126/science.aam7127,doi:10.1126/science.adh7691}. These results indicate that systems which are difficult to simulate with many-body methods probably host interesting physics.

The rest of the article is organized as follows. In Sec.~\ref{intro} we give a brief introduction of entanglement and average sign. In Sec.~\ref{result} we show the numerical results of the evolution of entanglement entropy, truncation error, and average sign with doping in the Hubbard model and discuss their correlation. In Sec.~\ref{discussion}, we give more discussions on the correlation between entanglement and average sign. We conclude our work in Sec.~\ref{Con_Per_Sec}. 
%%%%%%%%

%%%%%%%%
\section{Introduction of entanglement and average sign}
\label{intro}
To make this work self-contained, we first introduce the Hubbard model, then give a brief introduction of entanglement and average sign.

\subsection{Hubbard model}
The Hubbard model is arguably the simplest model of interacting fermions on a lattice but exhibits a wealth of correlated many-body physics. It is believed to be related to the microscopic mechanism of the high-Tc superconductivity in cuprate \cite{qinHubbardModelComputational2022,arovasHubbardModel2022}. The Hamiltonian of Hubbard model is
\begin{equation}
H = - t\sum\limits_{\left\langle {ij} \right\rangle \sigma } {c_{i\sigma }^\dag {c_{j\sigma }}} + U\sum\limits_{i } {n_{i\uparrow } {n_{i\downarrow }}} 
\label{Hubb_ham}
\end{equation}
where the first and second part are the kinetic energy $K$ and the on-site interaction $V$ respectively. The hopping constant $t$ is set as energy unit. The doping level is defined as $h = 1 - N_e/N$ where $N_e$ is the number of electrons in the system while $N$ is the lattice size.

\subsection{Entanglement Entropy}
\label{EE_TN}
Entanglement Entropy (EE) is a measure of the quantum correlations between different parts of a quantum state. By dividing the studied system into parts $A$ and $B$, a pure state $|\psi\rangle$ with density matrix $\rho = |\psi\rangle\langle\psi|$ can be written into a sum of product states: $|\psi\rangle = \sum_{i} \lambda_i |i_A\rangle|i_B\rangle$, where $|i_A\rangle$ and $|i_B\rangle$ are the Schmidt basis of the reduced density matrix $\rho_A$ and $\rho_B$, respectively. The von Neumann entanglement entropy can be expressed as:
\begin{equation}
    S_A = -\sum_i \lambda_i^2 \ln \lambda_i^2,
    \label{vNEE_eq2}
\end{equation}
which is the special case of the Renyi entanglement entropy:
\begin{equation}
    S_A^n = \frac{1}{{1 - n}}\ln ({\rm{Tr}}(\rho _A^n))
    \label{RenyiEE_eq}
\end{equation}
It can be shown that the von Neumann entanglement entropy is the Renyi entropy in the limit of $n =1$. 

In tensor network state related methods \cite{whiteDensityMatrixFormulation1992,schollwockDensitymatrixRenormalizationGroup2011,orusTensorNetworksComplex2019,RevModPhys.93.045003,annurev-conmatphys-040721-022705,xiang2023density}, 
the entanglement entropy of a quantum state determines the simulation accuracy with given resource.

\subsection{Average sign}
\label{Average sign}
We take the Auxiliary-Field Quantum Monte Carlo (AFQMC) \cite{AFQMC-lecture-notes-2013,assaadWorldlineDeterminantalQuantum2008} to introduce the definition of average sign. Similar definition can be also found in other QMC approaches, e.g., world-line QMC \cite{PhysRevB.26.5033}.  

For the Hubbard model defined in Eq. (\ref{Hubb_ham}), we can first perform the Trotter-Suzuki decomposition as
\begin{equation}
\begin{aligned}
    \exp ( - \delta (K + V)) = &\exp ( - \delta K/2)\exp ( - \delta V)\exp ( - \delta K/2) \\
    &+ O({\delta ^2})
\end{aligned}
\end{equation}
and Hubbard Stratonovich transformation in the spin decomposition form as
\begin{equation}
e^{-\Delta\tau U n_{i\uparrow}n_{i\downarrow}} = e^{-\Delta\tau U (n_{i\uparrow} + n_{i\downarrow})/2} \sum_{s_i=\pm 1} \frac{1}{2} e^{\gamma s_i (n_{i\uparrow} - n_{i\downarrow})}
    \label{spin_decom}
\end{equation}
where $\cosh (\gamma ) = \exp (\Delta \tau U/2)$
and $s_i$ is the introduced auxiliary field. We denote ${V_s} =  - \gamma \sum\limits_i {{s_i}({n_{i \uparrow }} - {n_{i \downarrow }})}  + \Delta \tau U\sum\limits_i {({n_{i \uparrow }} + {n_{i \downarrow }})/2}$.

According to Eq. (\ref{spin_decom}), the partition function $Z$ can be expressed as:
\begin{equation}
Z = {\rm{Tr}}({e^{ - \beta H}}) = \sum\limits_{\{ s\} } {{Z_s}}  = \sum\limits_{\{ s\} } {\det (1 + {B_s})} 
\end{equation}
where ${B_s} = \prod\limits_{m = 1}^{\beta /\Delta \tau } {{e^{ - K\Delta \tau /2}}{e^{ - {V_{s(m)}}}}{e^{ - K\Delta \tau /2}}} $.

In AFQMC, $Z_s$ is treated as the sampling probability. But there is no guarantee that $Z_s$ is always positive \cite{lohSignProblemNumerical1990a} (in some cases it can even be a complex number \cite{zhangQuantumMonteCarlo2003}). This is the origin of the negative sign problem. When the sign problem occurs, the absolute value of $Z_s$, i.e., $\left| {{Z_s}} \right|$ can be utilized as the sampling probability:
\begin{equation}
    {P_s} = \frac{{\left| {{Z_s}} \right|}}{{\sum\limits_s {\left| {{Z_s}} \right|} }}.
    \label{P_eq}
\end{equation}
The expectation value of operator $O$, i.e., $\left\langle O \right\rangle $ can be calculated as:
\begin{equation}
\begin{aligned}
   \left\langle O \right\rangle =\frac{{\sum\limits_s {{P_s}{\mathop{\rm sgn}} ({Z_s}){{\left\langle O \right\rangle }_s}} }}{{\left\langle {{\mathop{\rm sgn}} } \right\rangle }}
    \label{O_eq}
\end{aligned}
\end{equation}
where  $\left\langle O_s \right\rangle $ is the expectation value of the operator ${\hat O}$ for a single auxiliary field configuration and $\left\langle {{\mathop{\rm sgn}} } \right\rangle$ is the average sign:
\begin{equation}
\left\langle {{\mathop{\rm sgn}} } \right\rangle  = \frac{{\sum\limits_s {{Z_s}} }}{{\sum\limits_s {\left| {{Z_s}} \right|} }}. 
\label{avr_sgn_eq}
\end{equation}

For general system with negative sign problem, the average sign usually scale as $\langle \text{sign} \rangle \sim \exp(-c\beta N)$ where $\beta$ is the inverse of temperature, $N$ is the lattice size and $c$ is a constant \cite{lohSignProblemNumerical1990a}. From Eq.~(\ref{O_eq}), we can find that the observable $\left\langle O \right\rangle$ has huge fluctuation at low temperature in large systems. So the value of $\langle \text{sign} \rangle$ can be utilized as a measure of the difficulty in QMC calculation. Although it is known that there are cases where the negative sign problem is absent \cite{hirschTwodimensionalHubbardModel1985,PhysRevB.71.155115,PhysRevLett.116.250601,PhysRevLett.115.250601,PhysRevLett.117.267002}, general systems usually suffer from the negative sign problem.
%%%%%%%%

%%%%%%%%
\section{Results}
\label{result}
We study the Hubbard model on a $4 \times 16$ cylinder (with periodic (open) boundary conditions along the short (long) direction) with $U/t = 8$. 
We use AFQMC and DMRG to calculate the exact same systems.
We vary the filling factor and study a range of doping from zero doping (half-filling) to doping around $h=0.5$. We carefully examine the doping evolution of both the average sign in AFQMC at low temperature and the entanglement and truncation error in DMRG calculations of the ground state.   

\begin{figure}[htbp]
\includegraphics[width=0.48\textwidth]{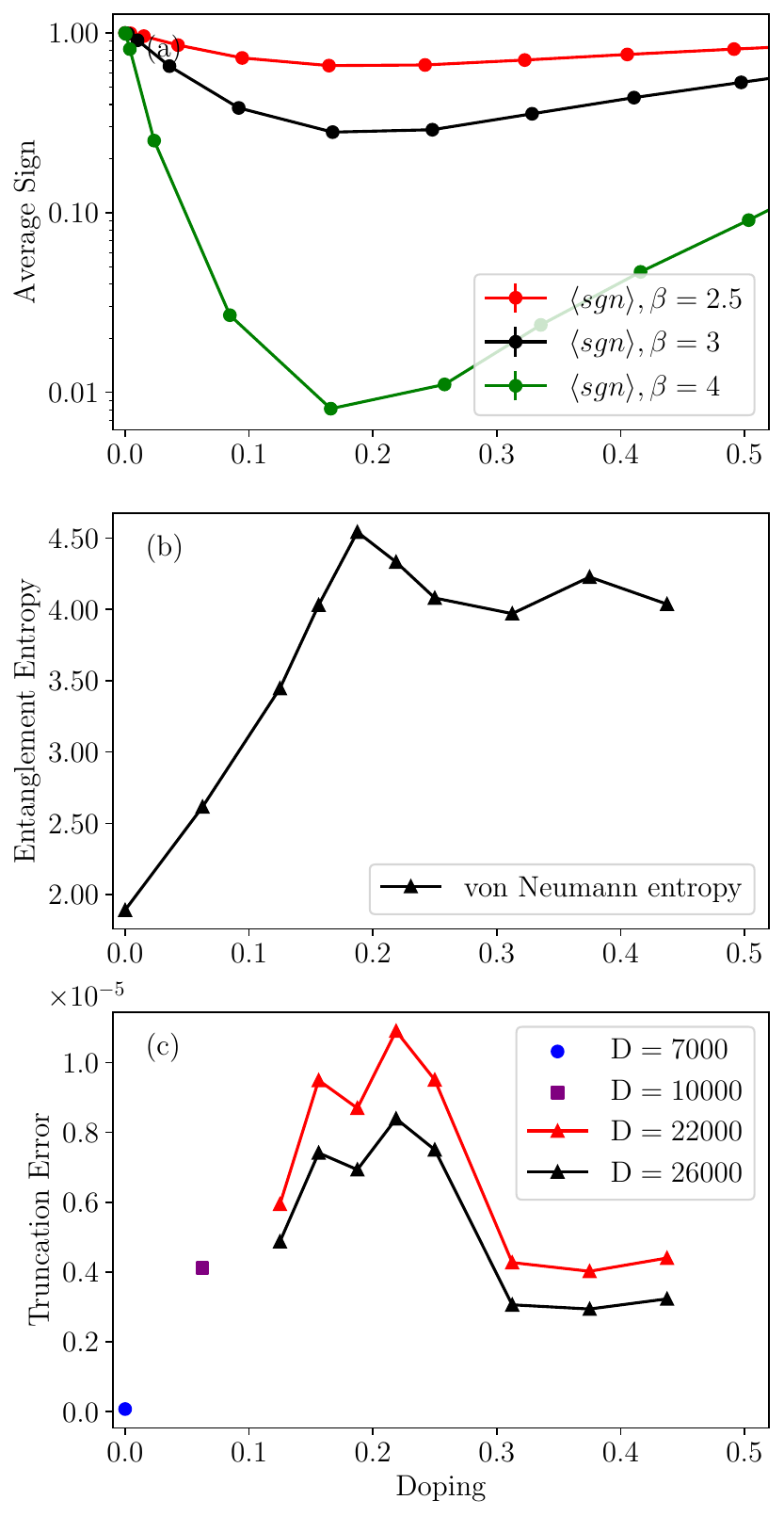}  
\caption {The evolution of (a) average sign in AFQMC, (b) entanglement entropy and (c) truncation error in DMRG with doping. In (a), average sign for $\beta = 2.5$, $3$ and $4$ are shown. In (c), the truncation errors in DMRG of different bond dimensions are shown. For the half-filling and the $h = 0.0625$ cases, small bond dimension already gives very small truncation error. We can find a correlation between average sign and entanglement entropy or truncation error. By doping away from the half-filling case, the average sign first decreases, reaching its minimum value at about $20\%$ doping, then starts to increase with the further increase of doping. The entanglement entropy shown in (b) are the values from a linear extrapolation with truncation error. The entanglement entropy and truncation error have opposite behavior as the average sign. Notice that region for small average sign and large entanglement/truncation error are difficult for both AFQMC and DMRG.}
\label{fig1}
\end{figure}

We show the evolution of average sign in AFQMC, entanglement entropy and truncation error in DMRG with doping in Fig.~\ref{fig1}. The truncation error in DMRG can be used to judge the accuracy of the calculation. In AFQMC calculation, we study two temperatures with $\beta = 2.5$ ,$3$ and $4$. Similar results can be found in many studies in the literature \cite{hirschTwodimensionalHubbardModel1985,lohSignProblemNumerical1990a,whiteNumericalStudyTwodimensional1989}. We can clearly see that the average sign decreases by lowing the temperature as expected. Af half-filling, the Hubbard model is sign problem free with $\langle \text{sign} \rangle = 1$ as shown in Fig.~\ref{fig1}. We find that the average sign first decreases when the system is doped away from half filling, reaching its minimum value around $20\%$ doping, and then starts to increase with doping. The entanglement entropy in Fig.~\ref{fig1} (b) are obtained from a linear extrapolation of the results with truncation error, whose values are plot in Fig.~\ref{fig1} (c) for different bond dimensions. For the half-filling and the $h = 0.0625$ cases,
small bond dimension already gives very small truncation error. The entanglement entropy and truncation error have opposite behavior as the average sign. Half filling system has the smallest entanglement entropy and smallest truncation error with fixed bond dimension.  With the increase of doping, the entanglement entropy and truncation error first increase and reach the maximum value also around $20\%$ doping, and then start to decrease. The results show that the region around $20\%$ doping is the most difficult region for both DMRG and AFQMC methods. Experimentally, $20\%$ doping is also the most intriguing region where different states close in energy intertwine with each other \cite{RevModPhys.87.457}. These
results indicate that systems which are difficult to simulate with many-body methods probably host interesting
physics. 

\begin{figure}[t]
    \includegraphics[width=0.45\textwidth]{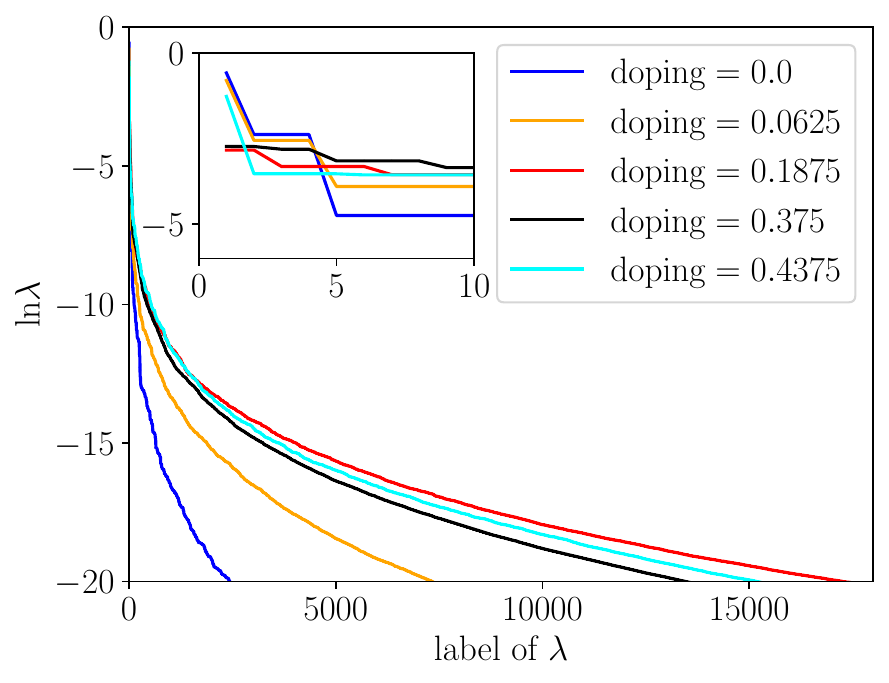}
    \caption{The entanglement spectrum for the Hubbard model on a $4 \times 16$ cylinder with $U/t = 8$. Results for doping $h = 0, 0.0625, 0.1875, 0.375$, and $0.4375$ are shown. The entanglement spectrum for $h=0.1875$ decay most slowly which is consistent to the large entanglement entropy and truncation error shown in Fig.~\ref{fig1}. The entanglement entropy at half-filling ($h = 0$) is steepest. The results for $h = 0.4375$ has a similar decay trend as $h=0.1875$ in the tail, but the leading entanglement spectrum has a steeper decay than the $h=0.1875$ case as shown in the inset.}
    \label{fig2}
\end{figure}

We notice that in panel (b) of Fig.~\ref{fig1}, the entanglement entropy has a bump around $h=0.375$. But the truncation error at $h=0.375$ is consistently small, meaning $h=0.375$ is consistently easier for DMRG calculation as other large dopings. To further examine the entanglement property, we plot the decay of entanglement spectrum in Fig.~\ref{fig2} where we show the entanglement spectrum for five typical dopings. We can find that the entanglement spectrum at half-filling (h = 0) is steepest, while the entanglement spectrum for $h = 0.1875$ decays most slowly, consistent with the entanglement entropy results in Fig.~\ref{fig1}. For large dopings, e.g., $h = 0.4375$, although the tail of entanglement spectrum is similar as $h = 0.1875$, the leading values decay faster than $h = 0.1875$, making the truncation error small in DMRG calculation as shown in Fig.~\ref{fig1}.    

We notice that the correlation between entanglement and average sign doesn't necessarily mean the system size can be handled by DMRG and QMC are the same. Although the vicinity of $20\%$ doping is the most difficult region for both methods, DMRG can accurately calculate the ground state of the studied $4 \times 16$ Hubbard model, while QMC have difficulty for the same system at very low temperature.

%%%%%%%%%%%%

%%%%%%%%%%%%
\section{discussion}
\label{discussion}

 Sign problem usually indicates the cancellation of positive and negative paths in the sampled space. The superposition of extensive positive and negative paths means massive destructive interference which usually results in highly entangled states. Physically, the competition between different candidate states in the low-energy manifold—whether due to frustration or the interplay between kinetic and potential energies—typically generates strong entanglement and a severe sign problem, leading to exotic quantum phenomena that are challenging to simulate classically. In this sense, average sign and entanglement entropy is naturally connected in challenging systems and can be used as an indicator to exotic physics \cite{mondainiQuantumCriticalPoints2022,PhysRevB.107.245144}.

It is known that the value of average sign depends on the chosen basis and details (e.g., the type of HS transformation) of the QMC approach. There exist methods to ease the sign problem by basis transformation \cite{nakamuraVanishingNegativesignProblem1998,hangleiterEasingMonteCarlo2020,levyMitigatingSignProblem2021}, but it is unlikely that the sign problem can be solved by local basis transformation for general systems, which is proved to be a NP hard problem \cite{troyerComputationalComplexityFundamental2005}. In \cite{okunishiSymmetryprotectedTopologicalOrder2014}, it is shown that the negative sign problem in 1D  bilinear-biquadratic spin-1 chain (including the Affleck-Kennedy-Lieb-Tasaki (AKLT) model) can be eliminated by the the non-local KT transformation and a local transformation. It is tempting to ask how general this conclusion is. For any given MPS, whether the corresponding frustration-free parent Hamiltonian \cite{2006quant.ph..8197P} can be transformed to negative sign problem free through some type of generalized KT transformation?

It is known that certain type of entanglement can be easily handled classically. For example, the stabilizer states formed solely by Clifford circuits are classical simulatable according the  Gottesman-Knill theorem \cite{gottesmanHeisenbergRepresentationQuantum1998}, though they can be highly entangled. In this sense, the Non-stabilizerness Entanglement Entropy (NsEE) \cite{qianAugmentingDensityMatrix2024,huangNonstabilizernessEntanglementEntropy2024} which is the entanglement entropy can't be removed by Clifford circuits determines the hardness of simulation with tensor network state related methods. It will be interesting to further investigate the possible correlation between NsEE and average sign. 

Our discussion so far has focused on the "exact" quantum Monte Carlo method. We notice that there also exist quantum Monte Carlo methods \cite{zhangConstrainedPathMonte1997,RevModPhys.73.33} where approximations are adopted to eliminate the negative sign problem. The essence of these methods is a trade-off between the negative sign problem and systematic error. Results has shown that the systematic error is mild in many cases \cite{PhysRevB.94.085103,leblancSolutionsTwoDimensionalHubbard2015}. It will be interesting to know whether there is a fundamental correlation between systematic error and entanglement since there is freedom in choosing the approximation to eliminate the negative sign problem.   

We also notice that although the solution of the negative sign problem is shown to be NP-hard \cite{troyerComputationalComplexityFundamental2005}, it is possible that we can gain the ``correct" physics by accurately studying large enough system size which is affordable in many-body simulations. The anti-ferromagnetic Heisenberg model on triangular lattice is an example, whose ground state is now known to have the $120$ degree order \cite{whiteNeelOrderSquare2007}. 

Our discussion in this work is for ground state or low temperature properties. It is known that both the average sign and entanglement entropy increases with the increase of temperature, which indicates that high temperature simulation becomes easier for QMC but harder for tensor network state related methods. However, we can take the purification scheme to make the tensor network simulation at high temperature easy \cite{verstraeteMatrixProductDensity2004}. 
%%%%%%%%

%%%%%%%%
\section{Conclusion and Perspective}
\label{Con_Per_Sec}
In this work, we study the correlation between entanglement and average sign, which measure the difficulty in the simulation of many-body system with tensor network state related methods and quantum Monte Carlo respectively. 
We study the doping evolution of entanglement in the ground state and average sign in QMC in the strongly correlated Hubbard model. We find they are indeed correlated. The entanglement entropy (average sign) shows a peak (dip) around $20\%$ doping, indicating that it is the most difficult region for both methods. It is also known that the vicinity of $20\%$ doping is the most intriguing region in both Hubbard model and cuprates where competing states with close energy intertwine with each other. These
results indicate that systems which are difficult to simulate with many-body methods probably host interesting
physics. Recognizing the correlation between entanglement and average sign provides new insight into our understanding of the difficulty of simulating strongly correlated quantum many-body systems. It will be interesting to also study the correlation between entanglement and average sign with the quantities which measure the difficulty in other types of many-body methods.

%%%%%%%%%%%%

\begin{acknowledgments}
The computation in this paper were run on the Siyuan-1 cluster supported by the Center for High Performance Computing at Shanghai Jiao Tong University. We acknowledges the support from the National Key Research and Development Program of MOST of China (2022YFA1405400), the National Natural Science Foundation of China (Grant Nos. 12274290, 12247103, 12204377), the Innovation Program for Quantum Science and Technology (2021ZD0301902), the Youth Innovation Team of Shaanxi Universities, and the sponsorship from Yangyang Development Fund.
%%%%%%%%%%%%
\end{acknowledgments}

\section*{Data Availability Statement}
The data that support the findings of this study are available from the corresponding author upon reasonable request.

%%%%%%%%%%%%Refs
\bibliography{ent}

\end{document}